\documentstyle[12pt]{article}
\begin{document}
\begin{center}

 THE DETERMINATION OF THE 'DIFFUSION COEFFICIENTS' AND \\
        THE STELLAR WIND VELOCITIES FOR X--RAY BINARIES.\\

        V.M.Lipunov $^{1,2}$ and S.B.Popov $^{1}$\\
 $^{1}$ Department of Physics, Moscow University, 119899, Moscow, Russia,\\
 $^{2}$ Sternberg Astronomical Institute,
 Universitetsky pr., 13, Moscow, Russia\\
 e-mail: lipunov@sai.msu.su and popov@sai.msu.su\\
\end{center}

 \begin{abstract}
 The distribution of neutron stars (NS's)
is determined by stationary solution of the Fokker--Planck equation.
In this work using the observed period changes for four systems:
Vela X--1, GX 301--2, Her X--1 and Cen X--3 we determined {\it D}, the
'diffusion coefficient',--parameter from the Fokker--Planck equation.
Using strong dependence of {\it D} on the velocity for Vela X--1 and GX 301--2,
systems accreting from a stellar wind, we determined the stellar wind velocity.
For different assumptions for a turbulent velocity we obtained
 $V=(660-1440)\, km\cdot s ^{-1}$.
 It is in good agreement with the stellar wind velocity determined by
 other methods.

 We also determined the specific characteristic time scales for
the 'diffusion processes' in X-ray pulsars. It is of order of 200 sec
for wind-fed pulsars and 1000-10000 sec for the disk accreting systems.

 \end{abstract}

\section{Introduction}
 The most precisely determined characteristic
for accreting neutron stars (NS's) is their period.
Thus using observations of the period we can determine
different properties of the observed object.

 Period changes show fluctuations. These fluctuations were discussed in
de Kool $\&$ Anzer (1993). The authors determined noise level in these systems
and characteristic time scales. Using these results we can estimate diffusion
coefficients in the Fokker--Planck equation (see below)
and the stellar wind velocities (only for the wind-accreting systems ).

 During accretion the angular momentum of plasma is transfered to the NS.
But the process of the momentum transfer is not stationary. The transfered
angular momentum fluctuates and therefore the period changes of the NS will
also show fluctuations.

 Processes with fluctuations are well known (see for example Haken (1978)).
Some applications
 of stochastic processes in astrophysics especially in accreting systems
were discussed in Lipunov (1987),Hoshino $\&$ Takeshima (1993)
 and Lipunov (1992).In Hoshino $\&$ Takeshima (1993) the authors,
 using simple models of MHD
turbulence, try to explain aperiodic changes in X--ray luminosity of
X--ray pulsars. Luminosity fluctuations are explained as the result of
density fluctuations due to turbulence in the plasma flow.
 The authors used 2D model for accretion disk and 3d model
 for the wind accreting
 systems. Detailed exploration of this question is very difficult in both:
 theoretical and observational ways (there is no good theory of MHD turbulence
 and the resolution of modern equipment of satellites is not high enough for
 power spectra of X--ray pulsars (see Hoshina $\&$ Takeshima 1993).
 But detailed exploration
 of the density fluctuations will help to understand period fluctuations. It
will be very interesting to compare X--ray luminosity fluctuations with
fluctuations of the period of the NS.

\begin{table*}
\caption[]{}
\begin{tabular}{|l||c|c|c|c|c|c|}
           & p, sec & p$_{orb}$, sec & L, erg/sec        & $\mu /10^{30}\,
Gs\cdot cm^3$ & t$_{su\, observ}$, yrs & t$_{su\, min}$, yrs\\
 Vela X--1 & $283$  & $7.7\cdot 10^5$& $1.5\cdot 10^{36}$& $3$  & $3000$&
$3000$\\
 GX 301--2 & $696$  & $3.6\cdot 10^6$& $10^{37}$         & $120$& $>100$& $100$
\\
 Her X--1  & $1.24$ & $1.5\cdot 10^5$& $10^{37}$         & $0.6$& $3\cdot
10^5$& $8000$\\
 Cen X--3  & $4.84$ & $1.8\cdot 10^5$& $5\cdot 10^{37}  $& $5.7$& $3400$&
$600$\\
\\
\end{tabular}
\end{table*}

 There are
different methods of describing of these processes. In this work we use
differential equations for the distribution function.
 For the frequency changes we can write the Langeven equation, which describes
the process with fluctuations:

 \begin{equation}
 \frac {d\omega}{dt}=F(\omega)+\Phi(t)
 \end{equation}
Here, $ F(\omega) $ --constant angular momentum. For $ F(\omega)$ in
the most general form we can write (Lipunov 1982):

\begin{verbatim}
F(\omega )\cdot I=\left\{
 \begin{array}
 & \dot{M}\eta _k \Omega R_G^2-k_t\frac {\mu ^2}{R_c^3},
 \qquad  {\text wind\, accretion}\\
 & \dot{M}\sqrt{GMR_d}-k_t\frac {\mu ^2}{R_c^3},
 \qquad {\text accretion\, disk},\\
 \end{array}
 \right.
\end{verbatim}
where $R_d$--disk radius, $k_t$ and $\eta _k$--dimensionless parameters
($k_t\approx 1,\, \eta _k\approx 1$), $I$--moment of inertia of the NS, $M$--
 mass of the NS, $\Omega$--orbital frequency of the system, $R_G$--
 radius of gravitational capture  ($R_G=\frac {2GM}{v^2+v_{orb}^2}$, we include
 here the orbital velocity) and $R_c$--corotational radius,
 $R_c=\left( \frac {GM}
 {\omega ^2}\right)^{1/3}$.

 We assume that the 'force' is conservative and in this case we can write
 $ F(\omega) $  in the form:
$ F(\omega)= -\nabla _{\omega}V $ , where $V$ is a scalar potential.
$ \Phi $  is a fluctuating moment, i.e. $ < \Phi(t) >=0 $ , (Lipunov 1987).

 The distribution of frequency, $ \omega $ , is described by the distribution
function $ f(\omega) $. This function satisfies the Fokker--Planck
equation (Haken 1978):

 \begin{equation}
  \frac { df}{ dt}= \frac { df \, F(\omega)}{ d\omega}+
   D \frac {d^2\, f } {d\omega^2},
 \end{equation}
where {\it D} is the 'diffusion coefficient', which is determined by the
correlation of the stochastic force $ \Phi $ :

 \begin{equation}
  < \Phi(t)\Phi(t') > = 2{\it D}\delta(t-t')
 \end{equation}

 Stationary solution of the eq.(2) is the following:

 \begin{equation}
 { f}(\omega)=N\exp {(-V(\omega)/{ D})}
 \end{equation}
where $N$ is determined from the normalisation condition:

 \begin{equation}
 \int_{-\infty}^{\infty} {f}(\omega) \, d\omega=1
 \end{equation}

 Using expression for $V(\omega)$ from Lipunov (1992)
 we can write {\it D} in the form:

 \begin{equation}
 { D}=\frac {k_t\mu^2}{3GMI}\cdot \frac{\omega^3}{\gamma}
 \end{equation}
where $ k_t $--constant parameter $ (k_t\approx 1) $ ,
      $I$--moment  of inertia  of the NS
and $ \gamma $ is evaluated as $ \gamma\approx \frac {t_{su}}{\Delta t} $ ,
here $ t_{su} $--time of spin-up and $ \Delta t $ --the characteristic time
for period changes (see for details  Lipunov (1987) or Lipunov (1992)).

 Using eq.(6) in 2.1 we shall determine the value of the 'diffusion
coefficient', {\it D}. With these {\it D} in 2.2 we shall make the
estimates of the stellar wind velocities for Vela X--1 and  GX 301--2.

\section{Results.}

\subsection{Determination of the 'diffusion coefficient'.}

 At first we shall estimate {\it D}, using eq. (6). In this equation
all variables, except $ \Delta t $ , are known (in principle). Their
values taken from Lipunov (1992) are shown in table 1.

 Nagase (1992) gives results of $GINGA$ observations of the cyclotron lines
in X--ray pulsars. For Vela X-1 and Her X-1 values of magnetic field, $B$,
coincides with the values $\mu$ that we used when radii of NS's
are $10\, km$ and $6\, km$ correspondently.
The value for $B$ obtained from observations for Vela X-1, $B=2.3\cdot
10^{12}\, Gs$, coincides quite well with the assumption that there is no
stable accretion disk in this pulsar.

 Characteristic time $ \Delta t $ for the wind--accreting systems can be
determined from the equation:

 \begin{equation}
 \Delta t \approx 1.7\cdot 10^{4} \alpha^{-2}10^{2(A+8.5)}L_{37}^
 {-\frac {12}{7}}\mu_{30}^{-\frac {4}{7}} sec,
 \end{equation}
where $ \alpha $ -- a fraction of the specific angular momentum of
the Kepler orbit at the magnitospheric radius, $A$--noise level (see table 2)
(de Kool $\&$ Anzer 1993).

\begin{table}
\caption[]{}
\begin{tabular}{|l||c|c|}
           &  A   & L$_{max}$  \\
Vela X--1  &$-9.1$& $10^{36.8}$\\
GX 301--2  &$-8.5$& $10^{37}  $\\
\\
\end{tabular}
\end{table}

 In eq.(6) for $t_{su}$ we must use minimum values. These times were
 calculated using equations for pure spin-up from Lipunov (1992),
 $t_{su\, min}$ are shown in table 1.

 We took $k_t=1/3, M=1.5\, M_{\odot} , I=10^{45}\, g\cdot cm^2.$
 From eq. (7) we can get
 $\Delta t$ for Vela X--1 and GX 301--2. For Her X--1 and Cen X--3,
 $\Delta t$ is determined from the graph in de Kool $\&$ Anzer (1993)
 (see table 3).
 So we can write equation for $D$ in the form:

 \begin{equation}
 {D}=5.55\cdot 10^{-19}\mu _{30}^2\omega ^3
 \gamma _6^{-1}I_{45}^{-1}\left(\frac M{1.5M_{\odot}}\right)^{-1}\, s^{-3} \, .
 \end{equation}

Values of {\it D} for four systems are shown in table 3.

 From the theory of diffusion we can write:

\begin{equation}
 D=\omega _{char} \dot{\omega} \, ,
\end{equation}

where $\omega_{char}$ is the characteristic length in the frequency space.

For characteristic time in this space we can write:
\begin{equation}
 t_{char}=\omega _{char}/\dot{\omega} =\frac {Dp^4}{4\pi ^2 \dot{p}^2}
\end{equation}

We can give a physical interpretation for $t_{char}$ for wind-fed pulsars
as a characteristic time of the momentum transfer:

\begin{equation}
  \frac {R_G}{v_{sw}}=400\left(\frac M{1.5M_\odot}\right)
  \left(\frac {v_{sw}}{10^8cm/s}\right)^{-3}\qquad  sec
\end{equation}

 For $v_{sw}=1178\, km/s$ we obtain $\frac{R_G}{v_{sw}}=245\, sec$ and for
$v_{sw}=1281\, km/s$ we obtain $\frac{R_G}{v_{sw}}=200\, sec$. These
velocities are close to $v_{sw}$ obtained using estimates of the
'diffusion coefficient'.

These characteristic time scales are also shown in table 3.

\begin{table*}
\caption[]{}
\begin{tabular}{|l|c|c|c|c|c|c|}
          &$\Delta t,\, sec$&$\gamma  $     &$D,\, sec^{-3}$    &$v_{sw},\,
km/s,\,
 $&$v_{sw},\, km/s,\, $&$t_{char},\, sec$\\
          &  &  &  & $(v_t=0.1a_s)$ & $(v_t=a_s)$ & \\
Vela X--1 &$1.5\cdot 10^4$  &$6.3\cdot 10^6$&$8.7\cdot 10^{-24}$&$848$ &$1442$
&$200$\\
GX 301--2 &$1.1\cdot 10^3$  &$2.9\cdot 10^6$&$2  \cdot 10^{-21}$&$656$ &$1120$
&$245$\\
Her X--1  &$4\cdot 10^3  $  &$6.0\cdot 10^7$&$4  \cdot 10^{-19}$&---   &---
&$960$\\
Cen X--3  &$2\cdot 10^4  $  &$9.5\cdot 10^5$&$4.1\cdot 10^{-17}$&---   &---
&$8500$\\
\\
\end{tabular}
\end{table*}

\subsection{Determination of the stellar wind velocity.}
 Fluctuating moment $\Phi $ (see eq.(1)) can be estimated as:
$$
\left(\frac {\dot{M}\, v_t\, R_t}{I}\right)
$$

 So for {\it D} we can write the equation which differs from eq.(6)
 (Lipunov \& Popov 1995):

 \begin{equation}
 {D}=\frac 12 \left(\frac {\dot{M}\, v_t\, R_t}{I}\right)^2\frac {R_G}{v_{sw}},
 \end{equation}
where $v_t$--turbulent velocity, $R_t$--characteristic scale of
the turbulence. This scale is of order of radius of gravitational
capture, $R_G$:

 \begin{equation}
 R_t\approx R_G=\frac {2GM}{v_{sw}^2}\approx 4\cdot 10^{10}v_8^{-2}\,
 \left(\frac M{1.5\, M_\odot}\right) cm,
 \end{equation}
where $ v_8=\frac v{10^8\, cm/s}.$

 The turbulent velocity is less or equal to the sound speed,$a_s$ (in opposite
case a great bulk of energy will dissipate in the form of shock waves),
that's why we can write:

 \begin{equation}
 v_t=\eta \cdot a_s,\qquad \eta \le 1
 \end{equation}
where $a_s=((5\, R\, T)/(3\, \mu ))^{1/2}=
 1.18\cdot 10^6T_4^{1/2}\mu ^{-1/2}\, cm/s $ and $ T_4= T/(10\,000 K)$.

 From eq. (12) we can get:

 \begin{equation}
 {D}=4.38\cdot 10^{-23}\dot{M} _{16}^2\eta ^2T_4\mu ^{-1}v_8^{-7}
 I_{45}^{-2}\left(\frac M{1.5\, M_\odot }\right)\, s^{-3}\, ,
 \end{equation}
 here $\dot{M} =\frac {\dot{M}}{10^{16}\, g/s},\ \ I_{45}=\frac I{10^{45}\,
 g\cdot cm^2}.$

 As we see there is a strong dependence of {\it D} on $v.$ So we can
evaluate $v$(in this case it is the stellar wind velocity, $v_{sw}$):

\begin{verbatim}
 \begin{equation}
 \begin{array}
 & v_{sw}=1700\cdot D_{-24}^{-1/7}\dot{M} _{16}^{2/7}\eta ^{2/7}T_4^{1/7}
 \mu ^{-1/7}\\
 & I_{45}^{-2/7}\left(\frac M{1.5\, M_\odot }\right)^{2/7} \, km/s.\\
 \end{array}
 \end{equation}
\end{verbatim}

 Values of $v_{sw}$ are shown in table 3.

\section{Conclusions}

 Obtained values of $v_{sw}$ coincide well with the characteristic
 value of this
 quantity for supergiants of early spectral classes: $v_{sw}\approx (600-3000)
\, km/s$ (de Jager 1980). We also use well known equation for a terminal
velosity of a stellar wind which is well confirmed by observations:

 \begin{equation}
 v_{\infty} \approx 3\cdot v_{esc},
 \end{equation}
where $v_{esc}=(\frac {2\, G\,M}{R_*^2})^{1/2}$ is the escaping velocity on the
surface of the star, $ r=R_*. $We use this dependence
and equation (de Jager 1980):

 \begin{equation}
 v(r)=V_{\infty} (1-\frac {R_*}r)^{\alpha}
 \end{equation}
where $\alpha \approx (0.35-0.5)$.

 For optical component of GX 301--2 we have: $M=35\,M_\odot , R=43\, R_\odot$
and $e=0.47$ (Watson et al 1982). We took $r=r_{min}\approx 2\, R_*,$
 because in our calculations we used maximum luminosity, i.e. luminosity in
periastr. For these values we have: $v_{esc}\approx 560\, km/s$ and for
$v_\infty$ we have $v_\infty \approx 1680\, km/s$ (for $\rho Leo$, B1 Iab, in
de Jager (1980) we can find $v_{sw}=1580\, km/s$).

 From eq. (18) we get: $v_{sw}(r=2\, R_*, \alpha =0.5)\approx 1180\, km/s.$
It coincides well with our maximum value $\approx 1150\, km/s$ (see table 3).
For Vela X--1 there are such estimates (Haberl 1991): $v_\infty =1700\, km/s,
 r=1.7\, R_*$ and $\alpha =0.35$. We get: $v_{sw}\approx 1250\, km/s.$ It also
coincides with our estimates of the maximum stellar wind velocity:
 $\approx 1450\, km/s.$ With different assumptions for the turbulent velocity
(for example $v_t=\frac 13 a_s$) we can get excellent coincidence of our
results
with the stellar wind velocities estimated above using other techniques.\\
{\it Acknowledgements.}\\
 We thank M.E.Prokhorov for helpful discussions and unknown referees
for there comments.

\end{document}